**Design Theory for Societal Digital Transformation: The Case of Digital Global Health**

Jørn Braa, HISP Centre, University of Oslo, Norway, jbraa@ifi.uio.no

Sundeep Sahay, University of Oslo, Norway, sundeeps@ifi.uio.no

Eric Monteiro, Norwegian University of Science and Technology (NTNU), eric.monteiro@ntnu.no

**Abstract**. With societal challenges, including but not limited to human development, equity, social justice, and climate change, societal-level digital transformation (SDT) is of imminent relevance and theoretical interest. While building on local-level efforts, societal-level transformation is a non-linear extension of the local level. Unfortunately, academic discourse on digital transformation has largely left SDT unaccounted for. Drawing on more than 25 years of intensive, interventionist research engagement with the digital transformation of public healthcare information management and delivery in more than 80 countries in the Global South, we contribute to theorizing SDT in the form of a design theory consisting of six interconnected design principles. These design principles articulate the interplay and tensions of accommodating over time increased diversity and flexibility in digital solutions, while connecting local, national, and regional/ global efforts at the same time.

Jørn Braa is professor at the University of Oslo, Norway, where he has spent much of his professional career initiating, cultivating and promoting interventionist research on and around the DHIS2/ HISP efforts. He is the "father" of the project. His work has been published in outlets including the MISQ, The Information Society, Information Technology for Development and in books.

Sundeep Sahay is professor at the University of Oslo, Norway and has the last couple of decades been engaged in the DHIS2/ HISP project, with a main focus on Asia in general and India in particular. His work has been published in outlets including the MISQ, Information and Organization, ISR and in books.

Eric Monteiro is professor at the Norwegian University of Science and Technology and an adjuct professor at the University of Oslo. He has a long-term, low-intensity engagement with DHIS2/ HISP over more than two decades. His work has been published in outlets including MISQ, EJIS, JAIS and one book.

# 1 Introduction

We analyze the efforts of societal-level digital transformation (SDT) to bring systemic change to public health information systems across multiple levels in and across countries. Drawing empirically from the Health Information Systems Program (HISP[1]), which has been ongoing since 1994 and organized around the design and implementation of an open source-based digital platform called District Health Information Software (DHIS), we examine the efforts aimed toward SDT in terms of improved equity, quality, effectiveness, and accountability in the provision of health services through improved information management. From its inception in three townships in post-apartheid South Africa, different versions and configurations of the DHIS are currently adopted by more than 80 countries, including all but two African countries south of the Sahara, Norway, and other "developed" countries, through the efforts of universities, Ministries of Health, international organizations, funding and various other agencies. Its ambition highlights the concerns of equity, inclusiveness, and ideology, that is, making design "that matters" (Bødker & Kyng, 2018) for society.

According to Markus and Rowe (2021), the study of SDT to date remains under-researched (cf. also Vial, 2019), despite its tremendous significance in the different facets of social life globally. This mirrors the general critique that IS research has privileged individual, group, organizational, and interorganizational units of analysis over a societal level (Faik et al., 2020). As digital initiatives permeate all aspects of social life, be they public health, education, humanitarian crisis, or consequences of climate change, there are both the potential and foreboding of how they will affect society (Walsham, 2012) that underscore the urgent need in IS research to thematize SDT.

---

[1] https://dhis2.org/

How, then, do we conceptualize SDT? SDT invariably implies *theorizing* the nature of societal-level dynamics based on but not reducible to local-level dynamics covered by digital transformation (DT) (Majchrzak et al., 2016). In our analysis, we conceptualize SDT as the interdependent interplay of local-level efforts to design and implement digital interventions in public health systems and how they shape and are shaped by societal-level changes across space and time. Furthermore, these local efforts are non-linearly aggregated across local/ district/national/ regional/global levels.

Specifically, our paper contributes to theorizing SDT in the form of a *design theory.* Our design theory is based on many local-level design interventions involving the evolving journey of DHIS. Responding to the criticism that the "evolution *across* projects [i.e., local interventions] is rarely considered" (vom Brocke et al. 2020, p. 521, emphasis in original) in design science research (DSR), our design theory thematizes the *relationship* between and across the many local interventions contributing to societal-level transformation.

The proposed design theory is formulated as an interconnected set of design principles (DPs). These DPs are presented in the form of a process model consisting of two interconnected levels: i) developing design interventions on the ground that build digital solutions for addressing local health information challenges and ii) scaling these local design solutions functionally and geographically to engage with larger societal concerns. The DPs articulate how SDT emerges from, but goes beyond, local interventions.

The rest of the paper is organized as follows. Section 2 reviews the relevant perspectives on DT, with an emphasis on arguments about the nature and mechanisms of *digital* transformation. We review and locate our design theory relative to other approaches within DSR. Section 3 presents the research methods, which are founded on theoretically sampled vignettes. Section 4 discusses four vignettes that demonstrate the characteristics of SDT.

Section 5 develops our design theory: the set of DPs that make up its core by problematizing their conditions and enabling circumstances. Section 6 provides the conclusion, which discusses the boundary conditions of our design theory and the conditions for its transferability to other contexts (including the Global North), and outlines future research.

## 2 Theoretical Framing

### 2.1 The D ("digital") in Digital Transformation (DT)

Some scholars (e.g., Vial, 2019) define DT generically as "processes" of technological change, while others narrowly define DT specifically to transform the "value proposition" or "organizational identity" (cf. Wessel et al., 2020; for a more comprehensive literature review, see e.g., Barthel, 2021, Baiyere et al., 2020, Reis et al., 2018, Vial, 2019). In sum, the definitions of DT oscillate between the unhelpfully broad and the overly narrow.

By contrast, we take our cue from Markus and Rowe (2021, p. 275): To attain DT, we need to identify the "digital properties that make a difference." That is, we need to understand the nature of digital technologies (Orlikowski & Iacono, 2001) to understand the qualifier "digital" in DT and SDT.

Zuboff (1988) coined the two terms *automate* and *informate* to capture the distinction between technology in general and digital technology in particular. In her analysis, all technologies, including digital ones, come with the potential to automate, reflecting the replicating of manual systems in digital form. What makes digital technologies distinct is their additional potential to also informate; that is, their outputs (data or information) may be recombined as inputs for generative purposes (cf. Henfridsson et al., 2018, Zittrain, 2006). Based on the processes of learning and empowerment, informate creates the potential for novel uses of digital technology, supplementing the intended ones captured by automation.

Informating recombines information in novel ways, in contrast to deskilling and disempowering, which are associated with automating.

Zuboff highlighted three sets of dilemmas: 1) the risk of deskilling and loss of meaning for workers though new representations of work; 2) knowledge affords power, which indicates that changes in knowledge may shift the existing power relations; and 3) informating aspects are not fully controllable by management, which causes unintended consequences by empowering workers.

The specific capacity of digital technologies to informate hinges on their ability to represent data indefinitely and without cost. Zuboff made the digital representation of the physical world a fundamental theme, while warning about the challenges of a digitally rendered "reality." Zuboff studied the safety-critical operation of running a large pulp mill. Changing from experience-based, embodied, and tactile handcraft—smelling, tasting, and feeling the temperature of the pulp—into a remotely operated, digitally enabled control room, she noted the unease stemming from "digital [representations] replacing a concrete reality" (ibid., p. 63) and "the sense of hands-on" (ibid., p. 65), which led to the "invent[ion of] ways to conquer the felt distance of the referential function." However, the representational capacity of data is exactly that–a capacity. It may, but most certainly need not, be actualized in the sense that data are woven into every day, consequential work practices (Günther et al., 2017; Mikalsen & Monteiro, 2021).

Zuboff's analysis "may have an even stronger story to tell now than it did when first published" (Burton-Jones, 2014, p. 72) as a result of the expanding scope and reach of datafication. The capacity to informate "renders events, objects, and processes...visible, knowable, and shareable in a new way" (ibid., p. 9). The representational quality of digital technologies highlighted by Zuboff should simultaneously be understood as an effort to push the limits of quantification into the realm of the qualitative (cf. Porter, 1996). Crucial to our domain of global health, one aspect of this involves methods of quantification to grasp the qualitative entity of "the whole" country or region. "Global," "national," or "regional" health are highly constructed entities. An immunization program manager in a developing country context needs to have data about the status of immunization of the entire catchment population he/she is responsible for to determine which children are excluded (Braa & Sahay, 2012). This gap between the normative requirement (of full immunization coverage) and the reality (of actual achievement) is quantified through indicators (percentage of fully immunized children) as a basis for action. Therefore, the key challenge is to identify, include, and—in our case—*digitally represen*t children not reached by health services.

Similarly, Didier (2020) provided a vivid ethnographic illustration of what went into the quantification of crop yields from "the whole of America." He analyzed the collective, distributed, and heterogenous practices—recruiting statisticians, editing and commenting on questionnaires returned from respondents, and refining the mathematical notions of random sampling and the sample frame—involved in constructing a unified measure of agricultural yields in the United States in the 1930s around the Great Depression. The relevance to our analysis is that digitalization quantification is key to achieving scale, but it occurs by devising routines, technology, and methods to *craft* a unified "whole." Zuboff underscored that what digital solutions "see" are only represented aspects of our lifeworld. In our case, this highlights how, when, and by whom the health status of the "whole" population becomes digitally represented.

## 2.2 Enter: The Societal Level

The literature on DT has "focus primarily on impacts located at the [local] organizational level" (Vial, 2019, p. 135); thus, attention to the societal level is lacking. This has led to calls for an expansion of DT to also address the societal[2] level, as suggested by Vial (2019, p. 136), who stated, "From this perspective, future research on DT may benefit from engaging with other streams of literature that focus on the higher-level implications of IT, including ICT4D (ICT for Development) as well as ICT4S (ICT for Society)."

However, shifting attention to the societal level implies more than merely changing the *empirical* unit of analysis, as it also implies an accompanying *theoretical* shift. Faik et al. (2020, p. 1360) pointed out that "the societal level reflects higher levels of complexity emanating from the multiplicity of rationalities that is characteristic of most modern societies." More forcefully, Majchrzak et al. (2016, p. 274, emphasis in the original) compellingly argued that a shift from an organizational to a societal level in IS research implies a simultaneous shift of theorizing that "replaces terms like *structure, decisions, antecedents,* and *control* by complexity theory inspired concepts of *emergence, catalyst, multi- level/multilayered, iterative, and conditions*…The concepts of processes and routines are forsaken; in their place are assumptions of continual evolution, localized tensions, and digital action repertoires." Building on this, we combine theorizing of emergence, flows, networks, and multilayered interconnection with due attention to specifically situated digital properties. We elaborate on some salient aspects as follows.

First, the above-identified characteristics of digital technologies of generatively allowing open-ended recombinations (i.e., repurposing) are consistent with the defining quality of public goods that are non-exclusive and non-rivalrous (Ostrom, 1990) as a "knowledge commons" (Frischmann et al. 2014). In our analysis, the key insight from a public goods perspective is that it functions as a (real) "social option." That is, it hedges against unforeseen future events and demand (Frischmann, 2012; cf. also Braa et al., 2004), generating a richer repertoire of future action possibilities. Investing in social options now "buys" increased downstream action possibilities and responses to unforeseen events, a key enabler for scaling to societal levels. At a technological level, this capacity for flexibility, is promoted by open-ended digital platforms with associated ecosystems (Tiwana, 2013).

---

[2] In the context of our design theory for SDT, we understand local as "one site of intervention" and societal level as "multiple sites of interventions, more often than not across countries".

Second, and as Castells (1996) argued, previously marginalized groups of people and geographical areas can seek to change their development status by becoming included in the "network society," and failure to do so risks further systematic marginalization. The inclusion in these informational networks leads to their marginalized condition becoming more visible. For example, Mosse and Sahay (2005) empirically analyzed how Niassa, a marginalized province in North Mozambique, sought to change their poor health status by joining the HISP network to make their health conditions more visible to health policymakers, creating an evidence base for more directed policy interventions and resources.

Third, we accommodate the need for SDT to account for inter-related, multi-scale dynamics, from the smallest health clinic to global policies and strategies. Across space, the biography (Williams & Pollock, 2012) or journey (vom Brocke et al., 2020) of digital technologies evolves as they are employed in new, local interventions. Across time, the evolution of digital technologies is inspired by what is understood in Ingold's (2013) perspective (cf. Baygi et al., 2021) as a flow of events with temporal intersecting modalities of timing (with new possibilities for action), attentionally (attunement to the exposure and absorption), and undergoing (ongoing becoming of the efforts). In our analysis, the multi-scale aspects of the case are clearly evident in the multi-level *interconnections* across space and time.

**2.3 Toward a Design Theory for Societal Digital Transformation (SDT)**

DSR follows a process of identifying a clear problem, building technology, intervening in the problem situation with that technology, and evaluating outcomes. DSR seeks to identify prescriptive principles or guidelines for the design of digital artifacts (Baskerville et al., 2018, Gregor et al., 2020, Hevner et al., 2008). Sein et al. (2011) formulated action design research (ADR) as a combination of DSR and the tenets of action research. They underscored the importance of designing for realistic organizational contexts. Resonating deeply with similar DPs in the user-centered design (Bødker & Kyng, 2018), interaction design (Svanæs & Gulliksen, 2008), or agile methods (Dingsøyr et al., 2019), ADR stresses how feedback from real-world organizational contexts becomes essential as the intended, but more often than not also the unintended, consequences of design materialize. However, our design theory for SDT deviates significantly from both DSR and ADR in three important ways.

First, consistent with a shift to a societal level, the clear goals of DSR and ADR that aim for "means-end relationships between problem and solution spaces" (vom Brocke et al., 2020, p. 2) are substituted with "emergence" and a "multi-layered" understanding of the problem (Majchrzak et al., 2016). Our conceptual approach emphasizes the long-term cultivation of opportunities in an improvised manner. In short, it is a design theory in which digital solutions "cannot truly be 'designed'" (Hanseth & Lyytinen, 2010, p. 4) but only cultivated, accounting for the force of history. This not only acknowledges the existence of unintended outcomes of design but also grants them a dominant role, underscoring the path dependencies (inertia) of prior design decisions. Nevertheless, our design theory recognizes two levels at which the local level corresponds more closely to DSR- and ADR-like aims. However, interventions at the local level subsequently trigger informating (thus unintended) aspects (cf. 'secondary design' in Gregor et al., 2020, p. 1631).

Second, DSR and ADR aim to enhance the quality and efficiency of the process and product (artifact) of design. This is different from the version of action research we subscribe to in which user participation is a political end in itself, essentially promoting democratic principles (Bødker & Kyng, 2018). To illustrate, the Iron and Metal Union Project (ibid.) in Norway at the shopfloor level sought to increase workers' and unions' influence when introducing digital technologies, such as numerical machine tools ("who should program them–workers or managers?"), through capacity building and empowerment in the area of digital technologies. At a societal level, this project lobbied (successfully) for changes in work–life legislation and union rights to negotiate when introducing digital technologies. By analogy, our design theory draws on an explicitly political agenda of enhancing citizens' and health workers' autonomy within global health while engaging in local design interventions. In the words of Bødker and Kyng (2018, p. 4:15), "[DSR] that matters should address changes that matter" in contrast to prevailing tendencies in DSR to "focus on small issues… such as products and technological solutions that the users like, rather than on solutions that profoundly change their activities as well as the goals they are supported in pursuing" (ibid., p. 4:2).

Third, the societal context is open-ended and thus not clearly defined and limited. The changes tied to SDT are, by necessity, global and also normative in aspiration. Accordingly, our design theory enables the processes of SDT to be spread—translated and replicated—across significant geographical areas in a self-sufficient, sustainable manner (Braa et al., 2004), thus responding to Majchrzak et al.'s (2016) call for conceptualizing the societal level as cutting across space and time.

## 3 Methods

The three authors of the paper are longstanding members of the HISP initiative and are engaged in various activities of empirical research in different countries, education, and advocacy. The nature, intensity, and focus of the three authors vary and have evolved over time. The first author is the "father" of the HISP initiative, who was directly engaged in setting it up in South Africa, and over time has been engaged in research, education, and engagement at the levels of policy and practice across nearly 40 countries over the last 25 years. The second author has been engaged in similar processes of engagement since 2000, with a greater empirical focus on the Asian region. The third author has also been involved since the mid-1990s but with lower intensity in terms of field-level engagement and greater involvement in research and supervision activities. This collective experience and knowledge of the authors has been developed based on the three modes of engagement elaborated below.

*Long-term lived experience in multiple contexts*: At least two of the authors of this paper have gone on field trips lasting months in different countries at different points in time, including South Africa, Mozambique, Ethiopia, Tanzania, Indonesia, India, and Vietnam. These visits involved primary data collection for particular studies, working with PhD and master's students, conducting system evaluation, teaching in partner universities, and building collaborations and institutional linkages with Ministries of Health, universities, non-governmental organizations (NGOs), and other civil society organizations. To illustrate this mode of engagement, in 2001, two of the authors visited Mozambique and established linkages with the public health and computer science departments of the national university and subsequently created tripartite agreements with Oslo, the national university, and the Ministry of Health. Based on this agreement, two key processes were set in motion. First, pilot studies of the HISP project were established in three districts in different provinces. Second, six students from our partner national university, three each from computer science and medicine, were selected for conducting their PhD studies registered in Oslo while working on the pilot projects in an action research mode. An interdisciplinary master's program in health informatics was established with Norwegian funding support, which remains active even today. Many political upheavals were experienced in the country, and the project was terminated in 2006. It was restarted in 2014, when the PhD graduates were back at home, occupying important positions in the Ministry of Health and in university. In 2019, with a new generation of PhD and master's students, the same two authors visited the original pilot districts to understand the processes of DT that had taken place over 25 years. This illustrates the nature of the long-term lived experience, spanning domains of research, education, policy, practice, and interventions with societal goals in mind. For example, when the project was initiated in 2001, Mozambique, still recovering from the ravages of the exodus of skilled staff with the exit of the Portuguese, followed by a long-term civil war, had no PhD graduates in computer science. Twenty years later, HISP was involved in producing eight PhD students.

*Engagement and supervision*: All three authors have been involved in the development of several research publications for journals and international conferences that have covered multiple issues of systems design and development, infrastructure challenges, concerns of sustainability and scalability of systems, technology transfer models, capacity-building strategies, and more. More than 50 students from the Global South have graduated with PhD degrees from Oslo, and the three authors have been extensively involved in their supervision, directly contributing to their own understanding of DT.

*Study of relevant research publications, master's and PhD theses*: The HISP research project and the use of the DHIS2 have become an object of research not only for various researchers within Oslo but also for many other scholars globally. It is interesting to note that despite the huge volume of publications that have emerged, none have used the label DT for their analysis but instead discussed related issues of organizational change, institutionalization, resilience, and sustainability. The analysis of these research outputs has been useful for us to understand not only what DT is but also what it is not and how the combination of various other concepts together and over time provides insights into the transformation phenomenon. For example, a seminal paper from the HISP project was on "networks of action" published in a special issue on action research (Braa et. al., 2004), which highlighted how networked action across and within countries sheds light on the challenges of sustainability and scalability of health information systems in the South. A study of this paper and putting it into a contemporary context helped to understand how the nature of such networked action has changed and how different strategies need to be cultivated to enable societal DT.

To capture the nature of this experience in a holistic manner, we relied on vignettes as a device for constructing a narrative of a phenomenon that has found extensive use in health research (Gourlay et. al., 2014). We deliberately selected four vignettes that reflect "revolutionary" moments in HISP history, highlighting notions of timing and attentionally (cf. Baygi et al., 2021), during the onset of the abolishment of apartheid in South Africa (1994), the spread of HISP to countries outside South Africa, the entry of the internet through fiber optic cables in Kenya, and the Covid-19 pandemic globally. During these moments of "punctuation" (Gersick, 1991), there tends to be a confluence of various activities that provides interesting insights into processes and mechanisms contributing to DT. Vignettes help to illustrate key practices or events that are relevant to a case study (Kotlarsky et al., 2014). They are presented in a narrative, story-like structure that preserves the chronological flow as opposed to what is normally limited to a brief time span, to one or a few key actors, or to a bounded space (Miles et al., 2013). In IS research, the use of vignettes has recently been adopted in some studies to interpretively develop a case narrative by presenting illustrative anecdotes from a larger story (Pozo & Sahay, 2017, Sæbo et al., 2021), reflecting particular themes the researchers want to highlight.

Our approach to the use of vignettes serves as an interpretive device to construct a narrative to theoretically illustrate DT at the societal level. This method is considered appropriate, as the overall empirical base is large and "messy," including different stories from multiple locations and actors over more than 25 years. Making sense of this story retrospectively would have been difficult through traditional modes of interviews and surveys, particularly because very few people would have had the holistic perspective of this complex and longitudinal story, which the authors cumulatively had to a larger extent. To build a narrative of societal DT, we recognize the need to start from the "whole" rather than standalone parts of the story. Thus, we agreed to build vignettes selectively sampled from different time points representing key moments in the DT process. A key theoretical assumption in this paper is that DT is made up of many incremental changes at multiple levels and time points, which cumulate, connect, and even disconnect to create transformational effects. Thus, having a longitudinal and historical perspective was key to building an understanding of societal DT. To determine which vignettes to include, the three authors jointly discussed different alternative vignettes that could best illustrate the narrative and their pros and cons using these criteria: i) they should illustrate different time points and geographical locations; ii) they should represent different levels of focus (i.e., global, national, health facility, and community); and iii) they should represent some unique aspects of DT, including the material features of technology involved, informed by our theoretical ambitions of building a design theory. The vignettes are not chronological in the sense that one follows strictly where the previous ends. Rather, they are partly pursued in parallel in HISP. Of course, there is danger in such an approach of creating selective bias, in which we only include the vignettes that confirm our theoretical conclusions and exclude those that may challenge or contradict it. We attempted to mitigate this risk by extensively discussing the pros and cons of what to include and exclude also with our fellow HISP researchers.

## 4 The Context and Evolution of Digital Transformation of Public Health Systems in the Global South

HISP is a global initiative that stands in (pleasant) contrast to big tech firms, such as Amazon, Facebook, and Google, and their surveillance capitalism (Zuboff, 2019). HISP seeks to improve the robustness and sustainability of the public health systems of countries in the Global South through research, education, and action toward "digital public goods" (cf.

Ostrom, 1990) or "knowledge commons" (Frischmann et al., 2014). HISP provides a unique blend of action research by combining research, education, and practical systems development and political advocacy for change.

The HISP research and development initiative is coordinated by the University of Oslo, Norway. It was initiated in three pilot districts in post-apartheid South Africa in 1994, with the explicit aim of supporting the newly decentralized health districts promoted by Mandela's African National Congress program by designing a health information system (HIS). The design approach was a combination of the Scandinavian democratic tradition of action research (cf. Section 2.3) with the South African grassroots anti-apartheid political movement. Over time, what started in South Africa now includes more than 80 developing countries that have adopted (with varying rhythms and intensities) the free and open source (FOSS) HISP digital platform called DHIS2 (District Health Information Software, version $2^3$). Never a stable design artifact, DHIS2 has developed and evolved through hundreds of design cycles. This development is coordinated by the University of Oslo and national implementations supported by country HISP groups (companies, NGOs, or university departments). This represents a transformative expansion over a period of 25 years, starting in South Africa in 1994.

Following Heraclitus, "No man ever steps in the same river twice, for it is not the same river and he is not the same man," the more than 25 years of the HISP journey have seen significant transformations in the world of public health, HISP's organizational and funding arrangements, the nature of the DHIS2 artifact, research and education in health informatics, and the authors' own world views. Only when these different changes at multiple levels and places are seen within a historical perspective can DT processes be discerned. In 1994, the HISP team comprised anti-apartheid activists, often seen as anarchists who constantly engaged in the battle against various governmental entities. Two decades later, HISP takes the contrasting position of the "mainstream," as it is funded and endorsed by a network of global partners (including the WHO,[4] PEPFAR,[5] GAVI,[6] UNICEF,[7] Norad,[8] and Global Fund[9]) and numerous national Ministries of Health that have adopted DHIS2 as the basis for their national HIS and supported by their sovereign budgets.[10] These changes in legitimacy, the nature and features of the artifact, data management processes, funding and implementation models, modes of political advocacy, and ideology have not come at once but in many small incremental steps forward and backwards, which shed light on the phenomenon of societal DT when interpreted cumulatively. There has been consciously cultivated collective action and learning across the network, supported by opportunistic events, with progress often coming from the failure of other competing projects and the endurance of HISP over time and space.

---

[3] https://dhis2.org/

## 4.1 Vignette 1: HISP in Post-Apartheid South Africa—Designing HIS for Decentralization and Strengthening Health Equities (1994–2001)

### 4.1.1 The context of the Problem Formulation

On the eve of apartheid, South Africa had one of the least equitable health care systems in the world. It served well the health needs of 20% of the (largely white) population but left the majority, mostly blacks, with dismal health services. During apartheid, there were 14 departments of health at the central level: the "general" Department of National Health, three for the apartheid-specific "White," "Asian," and "Colored" administrations, and 10 for the "black," "homelands," and "self-governing states." The different health services catering to different racial groups had limited coordination and non-existent data standards. By design, the bureaucratic organization promoted health inequity by obfuscating the capacity to compare health indicators across racial groups.

---

[4] The World Health Organization, https://www.who.int/
[5] President's emergency response to AIDS relief, https://www.hiv.gov/federal-response/pepfar-global-aids/pepfar
[6] Global vaccine alliance, https://www.gavi.org/
[7] United Nations' children's fund, https://www.unicef.org/
[8] The Norwegian agency for development collaboration, https://www.norad.no/en/front/
[9] The Global Fund to fight AIDS, Tuberculosis and Malaria, https://www.theglobalfund.org/en/
[10] Including: Tanzania, Mozambique, Nigeria, Bangladesh, Ethiopia, Ghana, Senegal, Democratic Republic of Congo, Kenya, and South Africa.

HISP was initiated with a base at the universities in Cape Town and Western Cape as a collaboration between anti-apartheid activists from the health sector and NGOs, university staff, and two Norwegian informatics PhD. students. The former PhD student (i.e., the first author of this paper) had a strong background in Scandinavian participatory design approaches, and the latter became the lead developer of the DHIS software based in South Africa for the next two decades and more. The first two years consisted of intensive work on (re)designing health data standards and formulating a roadmap for HIS development, consistent with African National Congress's Reconstruction and Development Program for the "new" South Africa, which sought to address the health inequities of apartheid by facilitating the comparison of health indicators and facilitating policy interventions (Braa & Hedberg, 2002). HISP received Norad funding in 1995 and was formally established in three health districts in Cape Town.

### 4.1.2 The Design, Building, and Intervention Processes

To understand and support the information needs of the district management teams, HISP's design work focused on three key areas: (a) development of digital HIS (called DHIS1) to initiate implementation in the three pilot districts; (b) populate the DHIS with "essential data sets" and standards for primary health care data, with the twin aims of reducing the data collection burden for health workers and improving the actionability of data for local decision making; and (c) creating a local team of super-users that could independently evolve the system design in the future based on changing informational needs (see Table 1 for a summary of vignette 1).

Given the explicit normative aim to design for equity, a priority was to collect standardized data from health facilities across racial groups to enable a comparative analysis. The general sentiment at the time, as expressed by the hospital manager in one of the pilot districts, was that "[a] lot of data is collected, but nothing is used." The widely recognized remedy was to focus on a minimum data set (MDS) of essential data to minimize the logistical challenge of collecting (often redundant) data from across health facilities. "Keep It Simple Stupid", the KISS principle, was seen as the way to go to create some order in the chaotic HIS field. The design challenge was to find a balance between making selected aspects of the health situation visible while acknowledging the efforts involved in doing so (Bowker & Star, 2000). Participatory design processes were employed to design the MDS along two trajectories: 1) negotiating with multiple stakeholders on what to include (and not) in the MDS and 2) participatory prototyping of the first version of the DHIS software based on the MDS (the first in 1997). DHIS was developed on the Microsoft Windows and Office platform, selected because it was the de facto standard in government at the time. The software was free and open source in all other respects. DHIS made it possible for districts to analyze their own health data for the first time. The user could compare one's own district data and performance with other districts. This ability was new and was a huge leap forward. According to the Eastern Cape information manager, "I can now eyeball data from all health facilities in the province and identify outliers and eventual problems–in a moment." This new possibility was empowering and created momentum that eventually led to the adoption of DHIS as the national standard in South Africa.

With no new funding in sight, a HISP Open Day conference in October 1998, which was planned as a wrap-up of the HISP pilot project, unintentionally turned out to be a trigger for subsequent scaling. Representatives from the neighboring Eastern Cape Province and the national level were present. Both were impressed, but in two different ways. Eastern Cape found the software solution attractive. Having defined their own version of an essential data set, they had failed to develop a software solution to effectively implement it. Conversely, the national level found the approach attractive. They struggled with a national MDS standard because of the difficulties in reaching agreement from all stakeholders. They were intrigued by how the approach in DHIS could resolve disagreements between provinces by giving them autonomy to include local data while simultaneously adhering to national data standards. This approach was conceptualized as the principle of the "hierarchy of information standards" (Braa & Sahay, 2013), and it subsequently became a foundation of the HISP approach. The MDS was crucial because it made it possible to cut through the red tape of organizational politics and a fragmented paper-based HIS and to agree on a subset of essential data items across different programs. As it was small and easy to collect, routine monthly data reporting from thousands of health facilities was made practically feasible.

An unintended consequence of Open Day was that HISP was 1) invited to implement DHIS in Eastern Cape and 2) to present the results in a national conference two weeks later. This triggered similar processes in other provinces, and a new national essential data set was gradually developed and formally adopted a year later. After obtaining an official endorsement in 1999, DHIS pilot projects, including the data sets, commenced in several provinces. In 2000, policies for the national rollout of DHIS in South Africa were developed, funded by the USAID[11] project that had invited HISP to Eastern Cape.

---

[11] The United States development agency, https://www.usaid.gov/

### 4.1.3 The Evaluation Processes

Evaluation was informal and consistent with the design and implementation processes being continuously validated by results on the ground and in a public setting, as exemplified by the Open Day event. Ultimately, it was the positive acceptance by users and the successful results on the ground in practice that secured continued support. In 2001, DHIS and HISP attained the status of a national HIS standard, and they have continued to evolve over the years, driven by users' demands for more. Today, South Africa is considered to have one of the most effective and functional HIS in the developing world (Braa & Sahay, 2012).

Table 1. Summary of Vignette 1.

| **Aspects of DT** | **Illustration** |
|---|---|
| Societal-level objectives | 1) Measuring health performance toward the political objective of increasing equity; 2) Decentralizing structures of governance to strengthen local empowerment |
| Design focus | Designing for MDS that enables minimizing health worker load and complexity in data collection while improving the comparability of health equity indicators |
| Design approach | Action research based on the principles of a bottom-up, user-led, evolutionary development |
| Technology/Software | Free and open source in spirit and practice, with a flexible metadata structure that enables agile development, user engagement, and evolutionary prototyping |
| Expanding local processes to build | Public demonstration of local successes creates momentum for a large-scale, low-cost national adoption that supports the political agenda of achieving health equity |

| societal-level alliances and impact | |

Design learnings from vignette 1 that feed into subsequently formulated DPs:

- Bottom-up, user-led design processes that cultivate locally useful solutions
- Flexible, modular solutions that promote Lego-like bricks for incremental transformation
- Minimum but essential data to minimize the workload of collecting data while allowing the comparison of health performances across sites inscribed in a flexible digital solution
- Mobilizing around the vision of equity through informating (thus empowering) local action

## 4.2 Vignette 2: From MDS to Data Warehousing—Action Research and Building the HISP network (2000–2010)

### 4.2.1 Context: Fragmented and Weak HIS as a Global Problem

The early problematization of the HIS in South Africa was not unique. Most developing countries exhibited similar challenges of fragmentation, lack of coordination, weak use of data and limited interoperability due to the largely manual nature of systems, with large amounts of unused data being collected (Sahay et al., 2017). In 2005, the WHO launched the Health Metric Network (HMN) as an initiative to strengthen the national HIS by prioritizing integration efforts. The HISP network effort from South Africa was expanding to other countries, including Mozambique, Tanzania, and Ethiopia in Africa and India and Vietnam in Asia. In this vignette, we describe these processes of scaling across and within countries and building research and educational capabilities to engage with the unique challenges of this scaling.

### 4.2.2 The Design, Building, and Intervention Processes

The South African experience galvanized HISP to expand its research and education capabilities by establishing university collaborations with master's and PhD programs in several countries, including Mozambique, Tanzania, Ethiopia, and, subsequently, Sri Lanka. These collaborations were built around practical DHIS implementation pilot projects, creating a unique blend of networks of action (Braa et al., 2004). In Mozambique, the informatics and medicine faculty from the national university,[12] in collaboration with the Ministry of Health, established three pilot districts across three provinces. Similar pilot sites were also developed in Tanzania and Ethiopia. These pilots exposed the systemic design limitations of DHIS1 due to its standalone architecture and the code being largely hardwired to South African needs. Data needed to be transmitted manually from districts to provinces using USB sticks or email attachments, representing a logistical nightmare. Moreover, the post-apartheid political ideology of equity in South Africa could not be replicated in other countries. Microsoft Access-based technology was outdated. As one evaluation report from Mozambique in 2003 stated, DHIS did "not allow for a professional three-tier system architecture." To respond to such criticism, the development of the web-based DHIS version 2 (called DHIS2) started at the University of Oslo in 2004, and working in collaboration with HISP India, the first implementation took place in Kerala, India, in early 2006. At that time, Kerala had a strong left-leaning political ideology, and the government welcomed an initiative based on FOSS.

These developments coincided with HMN's efforts to develop a national architecture framework built on the core concept of an integrated data warehouse as a key element. During a meeting with HMN in Geneva, HISP was asked if DHIS2 could function as a country's data warehouse, and they opportunistically answered positively. This led to an array of interconnected events, starting with the design and implementation of DHIS2 in Sierra Leone as the first country in Africa to follow a data warehouse approach. We learned that data warehousing was not merely a technical exercise; it implied building political coordination between different departments and donors and the different data systems they supported. The data warehouse design strategy, now better supported through the web-based DHIS2, enabled the integration and management of multiple data sets into one system. Data sets are typically represented by paper tools, register books, and reporting forms, and they are more complicated to change than digital tools. Therefore, the practical design strategy needed to address both the revision of the paper-based tools and the configuration of the data sets in the data warehouse by technically and institutionally addressing the conflicts between data sets (e.g., overlaps and incompatible definitions). Whereas some countries started the design with the existing data sets and reporting forms and solved the conflicts inside the data warehouse, such as Sierra Leone, others started with revising the paper-based data sets and then including them in the data warehouse, such as South Africa (Sæbø et al., 2021). Along with this cross-country movement of DHIS2, DHIS2 was gradually adopted by different states in India. These global and national processes increased the demand for constant improvements in the technical functionalities offered by DHIS2. The software rapidly evolved and became compatible with diverse health settings in varying political and institutional contexts.

---

[12] University Eduardo Mondlane

Working with HMN led to HISP's involvement in a WHO-led collaboration in establishing the "Open Toolkit," including interoperating open-source public good tools based on WHO standards[13] (Braa & Sahay, 2012). Along with DHIS2, an open medical record[14] and human resource software[15] formed the core of Open Toolkit, which was launched during a workshop in Ghana 2010, jointly organized by the WHO, the University of Oslo, and the West Africa Health Organization (WAHO). A proof-of-concept architecture of interoperability was established in a hospital in Sierra Leone and was demonstrated during the Ghana workshop.

---

[13] The standards are called SDMX, https://sdmx.org/
[14] Open Medical Record System (OpenMRS), https://www.open-emr.org/
[15] Integrated Human Resources Information Systems (iHRIS), https://www.ihris.org/

However, despite the high-level promotion by HMN, this interoperability architecture could not scale due to its inherent technical complexities. This taught a lesson on the limits of a top–down implementation of standards largely guided by technological imperatives in the Global South.

The Ghana workshop demonstrated the "successful" Sierra Leone DHIS2 experience to the other 15 member states in the region, prompting the WAHO health information to declare, "If DHIS2 can be implemented in [civil war-ravaged] Sierra Leone, it can be implemented anywhere." Following this meeting, WAHO and HISP established a formal collaboration, and an assessment of HIS was conducted in seven WAHO countries to develop a HIS-strengthening strategic plan. Two of the authors of this paper were directly involved in conducting the assessment. Sierra Leone was a frontrunner country, and the HMN had political motivations to declare the implementation there as a "success." However, the poor internet bandwidth in Africa limited the possibilities of the online implementation of DHIS2. Table 2 summarizes vignette 2.

**4.2.3 The Evaluation Processes**

A Ministry-supported evaluation from Mozambique concluded with a harsh verdict of the architectural deficiencies of DHIS1 and its use of Microsoft Access. This led HISP to initiate the development process of the web-based DHIS2. A second important evaluation was by HMN of the Sierra Leone implementation, which declared DHIS2 a "success" mostly for political reasons, triggering a wider uptake, such as in WAHO and Kenya. A formal evaluation of HISP and DHIS2 conducted by PATH[16] and commissioned by Norad provided more of a business analysis of DHIS2 and HISP in the global HIS landscape, highlighting the need to build the "ability of DHIS to align with national goals," and it "recommended prioritizing patient-level registration services in the core DHIS road map." The evaluation also came with a dire warning that if DHIS2 was not "moving in the direction of being a platform for broader registration of individual-level data services," it would not survive.[17]

---
[16] https://www.path.org/

Table 2. Summary of Vignette 2.

| Aspects of DT | Illustrations |
| --- | --- |
| Societal-level objectives | Addressing the challenges of technical and institutional fragmentation of HIS through improved integration |
| Design focus | Data warehousing through integrating and managing multiple data sets, not necessarily MDS but also large "raw" data sets |
| Design approach | User-led "learning pilots" supported by university collaborations, combining action, research, and education |
| Technology/Software | First full FOSS-based digital platform and relational database technology but did not have online web capabilities |
| Building societal-level alliances and impact | Raising awareness of digital solutions as a public good, shareable across and within countries, as a core strategy to address the core health system challenge of fragmentation. Research and education were vital for strengthening practical implementation efforts. |

Design learnings from vignette 2 that feed into subsequently formulated DPs:

- Collaborative, local action research fueling capacity building, partnerships, and alliances through trial-and-error design cycles with rich spillover learning
- Fostering technological and institutional integration by accommodating increased diversity in demand and requiring a data warehousing design approach from stakeholders

---

[17] https://path.azureedge.net/media/documents/DHS_norad_final_report_20h11.pdf

## 4.3. Vignette 3: 2011–2020: Mobile Internet, Cloud Computing, and Global Scaling

### 4.3.1 Context

As the problem of fragmented HIS was enduring, the HMN initiative raised political awareness of this challenge in more than 40 countries where HMN made master plans. The WHO health systems' strengthening framework (in 2005) explicitly placed HIS as a core component of systems strengthening. These global policy shifts provided an enabling environment for developing DHIS2 trials at the regional and national levels, as illustrated in the next vignette, separating the regional from the global level.

### 4.3.2 Regional Level: Mobile Internet arrives in Kenya and Africa

To demonstrate the value of the web version of DHIS2, an implementation site with sufficient bandwidth and a central server-based infrastructure was required. This coincided with the entry of the internet into Kenya. The previous HIS system in Kenya was based on the transfer of Excel files from districts to the national Ministry of Health. Aggregating health indicators was manual, thus cumbersome, and national overviews were not practically feasible. According to the Principal Secretary of Health at the start of the project, "We don't have data, and we don't know what happens in the districts across the country, and that is what we want to change with this new system." After establishing the first prototype, the team went to one hospital to test the feasibility of deploying DHIS2 online. However, after a power cut, the internet never came up again. Then, a volunteer pulled out a dongle connected to the mobile internet and suggested using it. What had happened was that an internet fiber optic cable had just reached East Africa and had been pulled up from the coastal town of Mombasa to Nairobi. This enabled DHIS2 to be implemented on a cloud-based server.  As an information officer from the Homa Bay district, a remote part of Western Kenya, stated, "I can access the data anytime and anywhere" and the principal secretary could say, "Now I have data from all over the country at my fingertips, even just by using my mobile phone." This allowed for the comparison of his district data with those of neighboring districts. As in South Africa, DT contributed to empowering local-level information managers by informating their work environment and adding value to their work in relation to management and others. The results in Kenya opened the doors for HISP, first in Uganda and Rwanda in East Africa and then in Ghana and Burkina Faso in West Africa. Thereafter, all but one of the 16 WAHO countries followed suit. The East Africa community followed in creating a regional database. Inspired by the technical advancements in DHIS2 and its increasing acceptance in Africa, Mozambique, which had discontinued DHIS in 2005, reinstated DHIS2 and made it its national system in 2015. The local team, fresh from their studies in Oslo and drawing upon the graduates of their national master's program, began to technically support Portuguese-speaking WAHO countries, such as Cape Verde and Guinea Bissau.

### 4.3.3 Global Level: WHO releases Digital health Packages for DHIS2—Enabling Diversity and Flexibility

In 2015, the WHO and HISP started a collaboration on developing health program-specific digital health packages, such as for tuberculosis, immunization, HIV/AIDS, and malaria, including standardized metadata, which could be downloaded and easily implemented in DHIS2 (Poppe et al., 2021). This, together with the further development of the Open API, effectively transformed DHIS2 into an evolving platform ecosystem that allows for an open-ended set of differentiated services. These packages were developed through the advancement of the so-called DHIS2 Tracker app for case-based, individual, and patient data. During the Ebola crisis 2014-2015, the HISP support to Liberia led to a "peak moment in Tracker development" (core DHIS2 developer) where the new relationship model enabling contact tracing represented a significant undertaking. As the process of developing the digital health packages gained momentum, in 2018, guidelines for data analysis and curricula for training were also included in the packages in collaboration with various WHO health program departments. These packages represented a new digital and pluralistic take on the essential data set design approach as an effort to focus on need-to-know data in each health program and as part of health services but within the same system.

Another major impetus for the global scaling of DHIS2 came from its adoption by global donors. In 2015, while PEPFAR was looking for a new system for reporting HIV data from multiple contractors in more than 50 countries, it realized that DHIS2 was already the national system in most of these countries. Thus, they selected DHIS2 as their new system because it could facilitate integration with national systems in the future. With PEPFAR leading the way and with increased legitimacy through the WHO collaboration (University of Oslo was made a collaborating center for WHO for HIS strengthening), the Global Fund, GAVI, and other donors joined forces because they were also receiving requests from countries to support DHIS2. They saw the practical value of developing a funding model for DHIS2 as a means toward global public good. New knowledge, particularly on the digital side, was provided by University of Oslo, which, by 2017, had a large group (about 70 staff, including those based remotely) exclusively supporting software development and implementation in countries, conducting advocacy for funding support, and enabling research and education. In 2020, the Global Fund promoted the establishment of regional HISP nodes, where HISP country teams (e.g., India, Bangladesh, Sri Lanka, and Vietnam) would directly get funded as "HISP South Asia" to jointly support more than 10 countries in the region. Table 3 summarizes vignette 3.

**4.3.4 The Evaluation Processes**

This vignette covered a period of 10 years and spanned many countries. It involved various formal evaluations, notably by the funding agencies, and was linked to their concrete funding supplemented by various ongoing informal evaluations. An evaluation of HIS strengthening strategies in seven WAHO countries by two University of Oslo researchers was supported by USAID. Country-based evaluations were also conducted, such as in Ethiopia, Sri Lanka, and Bhutan.

In 2016, another high-level formal evaluation of the HISP and DHIS2 network was instituted by global donors and agencies, funded by Norad, and conducted by the same PATH agency as in 2011.[18] Both of these evaluations provided important feedback crucial for establishing a level of legitimacy, a funding structure, and accountability. The 2011 evaluation (see the previous vignette) led to an increased focus on individual patient-level data, while the 2016 report highlighted the value of the HISP research and education network in fostering digital innovations.

Table 3. Summary of Vignette 3.

| Aspects of DT | Illustrations |
|---|---|
| Societal-level objectives | Capturing the "whole" and engaging with the problem of invisibility of data, particularly marginalized ones, by leveraging on the opportunity provided by the arrival of the internet in Africa |
| Design focus | Addressing increased diversity in demand through online data warehousing to obtain full-coverage data and to develop multiple specialized data sets (e.g., malaria, HIV, and tuberculosis, vaccination) |

---

[18] https://path.azureedge.net/media/documents/DHS_HISP_assessment_final_report.pdf. It should be noted that these evaluations from PATH were largely on systems-related processes and were not exclusively for research processes. This part of the evaluation could come from different studies and theses of researchers within the broader HISP community.

| Design approach | Building a country DHIS2 core team through hands-on learning in evolutionary development (with the country being "bottom" for creating local ownership in a global context); leveraging on multi-level global, regional, and national connections through the use of WHO apps |
|---|---|
| Technology/Software | Online cloud-based data warehousing. Open API and platform capabilities with interacting apps across both aggregated and case-based data systems |
| Building societal-level alliances and impact | Practically advancing the normative agenda of digital public goods for health though multi-level- (national, regional, and global) and multi-sectoral- (ministry, universities, donors, and NGO) based collaborative efforts addressing a diverse range of health areas |

Design learnings from vignette 3 that feed into subsequently formulated DPs:

- Capturing "whole" populations by leveraging mobile internet and online data warehousing, improvising on the arrival of a new high-speed connection to Africa
- Accommodating increased scale and diversity through technical and institutional federated organizations and semi-autonomous local teams for capacity building and online warehousing technology, together with new functionality/modules
- Capacity scaling through university-led research network building coupled with partner alliances, both local and global

**4.4 Vignette 4: Covid-19 Platform—Increased Diversity through Interoperability (2020 – ongoing)**

**4.4.1 Context: The Crisis of the Covid-19 Pandemic**

In early 2020, the Covid-19 pandemic was characterized by uncertainty and rapid changes in the disease situation, policies, and practical measures in responding to the pandemic. Control at border points, lockdowns, testing and tracing of positive cases and their contacts, linking to laboratories, and monitoring of intensive care units were among the multiple measures conducted during the first part of the pandemic, all of which had specific requirements for informational and digital solutions.

In early 2021, Covid-19 vaccination became the key pandemic response, along with information and digital responses, scheduling and monitoring in many countries, limited vaccine stocks, QR-coded Covid-19 certificates, and adverse effects of vaccination. Data on Covid-19 cases, deaths, and hospitalizations, as well as data on vaccination and its impact, have been collected and managed, closely monitored, analyzed, and acted upon in all countries, resulting in, or used to legitimate, a variety of pandemic responses, with high societal impacts. One consequence of this level of emergency was that governance and decision-making powers related to designing and implementing new digital solutions were shifted to a high governmental level, thus heightening political visibility.

### 4.4.2 The Design, Building, and Intervention Processes

Sri Lanka was the first country to use DHIS2 to design digital Covid-19 responses, starting already at the onset of the pandemic in January 2020, and develop solutions that were disseminated, further extended, and applied globally. Sri Lanka could take a leading role in this development because the country had historically used DHIS2 as its main HIS and knew it well, had been running a health informatics master's program since 2010, and had institutionalized the cadre of Information Officers within the Ministry of Health manned by graduates of the master's program. Further, a strong HISP Sri Lanka group was developed, with a strong grounding in research, education, and practice and reliable expertise in DHIS2.

The design of the first DHIS2 Covid-19 app in Sri Lanka was based on the workflow of the initial registration of arrivals at the port of entry. Testing and follow-up visits in the community were presented, approved, and deployed rapidly just prior to the country reporting its first coronavirus case on January 27, 2020. Following the first indigenous case in the second week of March 2020, the Ministry shifted its focus to the clinical management information of confirmed cases, including data on sociodemographic factors, symptoms, daily clinical updates, laboratory information, and outcomes. The HISP team could rapidly repurpose this functionality using the DHIS2 Tracker module based on their prior competence and later extend this to also include suspected case tracking (see Figure 1 for a timeline). New requirements were related to the automation of transmission of information, links to other systems, and certain required visualizations, such as contact mapping. To effectively deal with these rapidly increasing new informational requirements, a hackathon was organized by the Sri Lankan Government's ICT Agency[19]. This attracted volunteer developers from around the world, representing public–private players, and agile development methods were applied. The hackathon helped create various outputs, such as the contact tracing web application, intensive care unit bed management application, tracing location through mobile towers, and integration functionality with the immigration department system. This contact tracing app was made possible by applying the relationship model developed in the DHIS2 during the Ebola outbreak in Liberia five years earlier. Another product of the hackathon was the myHealth mobile application, which enabled public messaging and alerting citizens of potential contacts with infected cases. Eight new modules were developed in a short span of three months (see timeline below) during the peak of the first phase of the pandemic.

---

[19] https://www.icta.lk

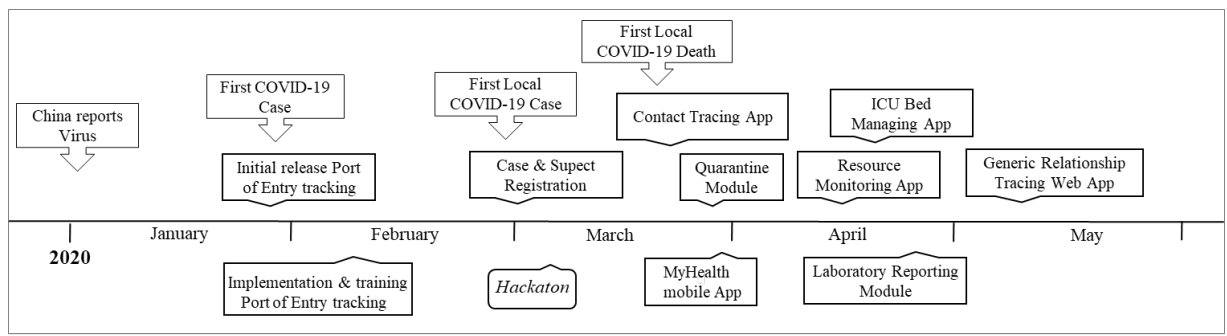

Figure 1. Timeline of the Development of new Modules in Sri Lanka (Amarakoon et al., 2020)

A developer commented on the agility of the process:

> "*it usually takes weeks or even months to get approval and implement a system within the ministry. But during the early days of the pandemic, it was hard to believe how stakeholders outside the ministry and sometimes even outside the country could collaborate with ministry officials and get systems designed and implemented*".

The DHIS2 core team in Oslo also helped create changes in the DHIS2 core where needed and engaged in rendering the apps and modules generic for further global dissemination. This unique mix of people and modes of digital collaboration on the DHIS2 platform, combined with fast, high-level decisions on the deployment of new systems, helped develop rapid innovations in weeks, which otherwise would have taken years to build.

The variety of apps and modules designed and deployed using the DHIS2 platform, as well as interoperability with other systems, such as for immigration, laboratories, and population registers, made up the information and digital responses to the pandemic in Sri Lanka. The emergence of the pandemic led to the rapid development of a whole ecosystem of need-based interoperating systems based on the situation on the ground in terms of existing systems, capacity, and governance structures. This design process can be called bottom-up "architecting," in contrast to the earlier efforts made during the HMN process to establish an architecture top-down. The contact tracing and surveillance app and other modules developed and used in Sri Lanka were then further developed through global collaboration and participation by the core DHIS2 team into a more generic "digital package" made available for global distribution in March 2020.

As soon as it was released, HISP Mozambique obtained the Covid-19 package, translated it into Portuguese, adapted it to local needs, and implemented it in Mozambique. They also developed new modules, such as data interoperability with laboratories. Moreover, HISP Mozambique supported the Portuguese-speaking countries of Cape Verde, Guinea-Bissau, and Sao Tomé in implementing Covid-19 modules. Recently, QR codes for vaccination and digital certificates and modules for self-registration for Covid-19 vaccination have been developed by HISP Vietnam for implementation in Laos. This was adapted by HISP Mozambique in its context. More than 40 countries have adapted DHIS2 modules for Covid-19 mass vaccination.

Norway, learning from Sri Lanka, implemented DHIS2 for contact tracing in municipalities[20] in 2020. The medical head in one municipality said, "The policy and requirements of the pandemic change all the time. The good thing with DHIS2 is that we can carry out changes to the system to meet new requirements ourselves. We don't need to wait for anybody." In Norway, the emergence of the pandemic took data sharing to levels not witnessed before. Initially, each municipality ran its own contact tracing system, as it did with other disease surveillance registers. However, to trace contacts and positive cases across municipal borders, they quickly agreed to share data across municipalities through a national inter-municipality portal. Access to the Norwegian population register, which also allowed citizens' online access, and to the national lab system were quickly established, thus providing an entire ecosystem of interoperating systems. Norway's adoption of DHIS2 represented a unique case of a "reverse innovation," challenging dominant technology transfer models of innovations flowing from the North to the South (Sahay et al., 2017). Table 4 summarizes vignette 4.

---

[20] Note that this DHIS2 app is not the same as the so-called Smittestopp ("Stop the spread") app developed under the responsibility of the Norwegian Institute of Public Health. The Smittestopp app was launched quickly but later had to be stopped due to privacy issues.

### 4.4.3 The Evaluation Processes

The Covid-19 pandemic is still unfolding, and there have been no formal evaluations of the role of DHIS2 in providing digital responses. However, DHIS2 has been assessed and recommended as a leading digital solution for Covid-19 response by institutions, such as Johns Hopkins University and the Center for Disease Control (CDC): "[T]he platforms chosen for review are mature platforms that have been deployed all over the world." In the context of the Covid-19 use case involving patient triage, referral for testing, and contact listing and follow-up, the two platforms that stand out for their maturity, flexibility, and large-scale deployment are DHIS2 Tracker[21] and CommCare. The CDC's Guide to Global Digital Tools for Covid-19 Response compares and recommends DHIS2 and other digital tools, such as CommCare.[22]

Table 4. Summary of Vignette 4.

| Aspects of DT | Illustrations |
|---|---|
| Societal-level objectives | Providing digital support to the societal challenge of Covid-19, which endangers the health and lives of whole populations |
| Design focus | Building multiple informational response-based interoperable platform ecosystems and app development |
| Design approach | Collaboration on design, implementation and innovation in the global FOSS network of developers, implementers, and users |

---

[21] https://drive.google.com/file/d/1yCP7t1di_ofQ0YhuPAD1Oqcj1aTo74k5/view
[22] https://www.cdc.gov/coronavirus/2019-ncov/global-Covid-19/compare-digital-tools.html

| Technology/Software | Interoperable platform eco-systems with apps, modules and data sharing with multiple systems. API and Tracker further improved |
|---|---|
| Building societal-level alliances and impact | DHIS2 Covid-19 apps were first developed in Sri Lanka and further enhanced in the global network, promoted through WHO and GAVI, spread to more than 40 countries including Norway |

Design learnings of vignette 4 that feed into the subsequently formulated DPs

- Promoting an evolving platform architecture to enable an ecosystem of modules, functions, and apps, addressing a diverse and rapidly changing demand responses to a health crisis
- Promoting new forms and modes of innovation networks spanning the Global South and North, thus dismantling the center versus periphery of technological innovation
- Long-term investment in capacity building to enable rapid repurposing of new demand as a social option

## 5 Analysis

In this analysis section, we provide a synthesis of the design learnings distilled from the four vignettes (Section 5.1) and describe the proposed DPs inferred through this synthesis (Section 5.2). In Section 5.3, we elaborate on the theoretical and in Section 5.4 the practical implications of our theory for SDT.

### 5.1 Synthesis of Learnings

We begin by schematically representing the synthesis, which is illustrated in Figure 2 at two interconnected levels. One level concerns the horizontal dimension (or axis) representing the temporal dimension depicting the local-level design interventions and innovations that have taken place in different country contexts. We conceptualize these developments over time to highlight the increasing diversity of digital systems, which is a result of flexibility in the design provided for increased specialization and thus granularity. The vertical dimension (or axis), which metaphorically represents a spatial dimension, highlights the mechanisms through which local-level digital developments become woven into transformations at the societal level within and across countries, with global consequences. This link between local-level digital innovations and their translation into societal-level values provides the core theoretical proposition of our theory for SDT.

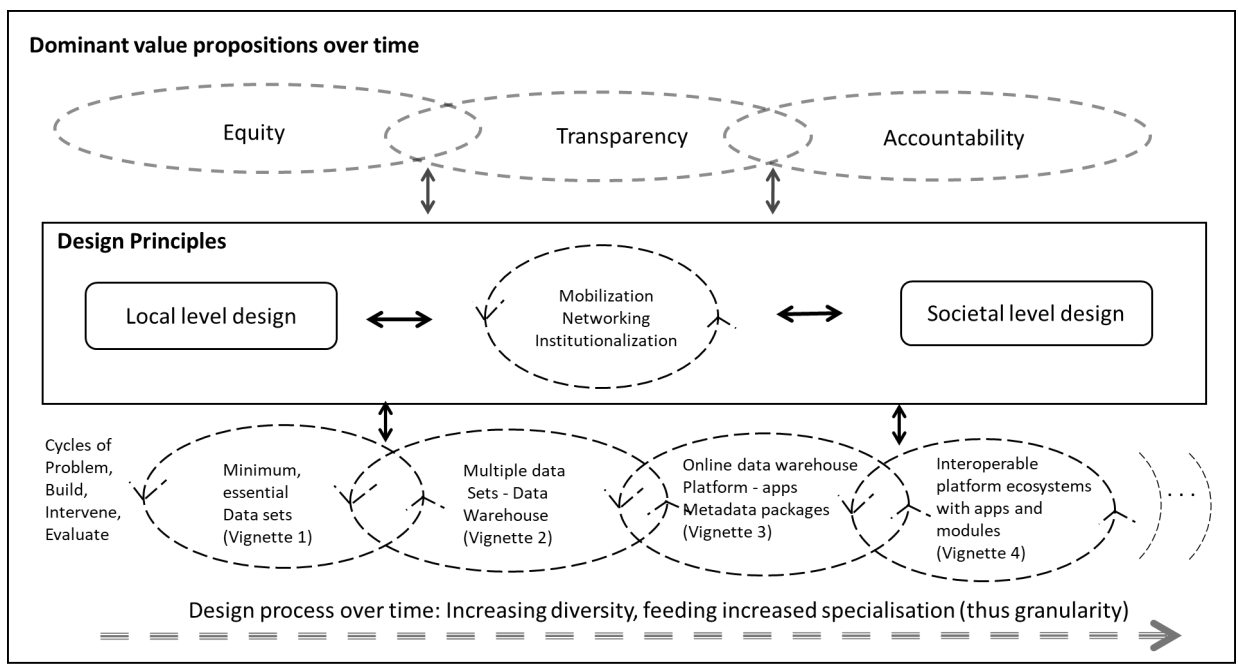

Figure 2. The Proposed Model for Societal Digital Transformation (SDT) with its Constitutive Design Principles (DPs).

The four vignettes represent unique moments from HISP history, illustrating the localized design interventions and innovations shaped by contextual conditions, infrastructure, capacity, and health system priorities. Over time, we see how these innovations have expanded in scope to accommodate the increased diversity of demands. This represents the core principle of the synthesis of our design learnings. In South Africa, the focus was on rebuilding an apartheid-ravaged health system characterized by extreme fragmentation and centralization. Responding to this was the design of an essential data set and its incorporation into the DHIS software. Beyond a technical solution, also it represented a negotiated balance of practical feasibility and political advocacy for equity. In the second vignette, there was a move toward multiple (unlike the focus in South Africa on singular) data sets within a data warehouse framework in response to the increased diversity in demand. This motivated the development of DHIS2 based on the core principle of a flexible metadata structure from DHIS1 but expanded to incorporate the principles of a web-based data warehouse capable of extracting data from different data sources and managing multiple data sets. In contrast to this stand-alone architecture, the next vignette saw the cloud-based data warehouse scaling countrywide, first in Kenya and then across Africa. This made data "visible" from remote and marginalized parts of the country (Zuboff, 1988). The subsequent development of an Open API enabled a platform architecture that formed the basis for WHO-distributed digital health packages. The fourth vignette, born during the Covid-19 pandemic, motivated the evolution of DHIS2 into a full-blown interoperable platform ecosystem architecture with associated apps and modules developed by the HISP network of associates and partners, including institutions (e.g., immigration) outside of health.

In summary, the development of design interventions in different local contexts fueled the capacity to accommodate an increased diversity of demand over time. More granular (i.e., specialized) health reporting was enabled through the WHO's endorsement of the DHIS2 for its disease-specific health programs (e.g., HIV/ AIDS, tuberculosis, and malaria). Thus, increasing diversity in demand was not only incorporated as different data sets but also as apps and modules addressing specific needs of different health programs, indicating increasing specialization.

Along the vertical dimension in Figure 2, we represent how local-level, country-specific design interventions were translated into transformations at the societal level. Although these societal-level implications were first specific to the local country, they were then gradually also made visible across countries. In South Africa, there was the explicit normative aim of advancing health equities, and digital systems were designed to support this aim, such as by allowing inter- and intra-district comparisons to identify which were more disadvantaged and to design policy interventions accordingly. There were many mechanisms through which local, district-specific innovations were translated to the national and societal levels, including political advocacy and the demonstration of results in provincial and national seminars. For example, in Sierra Leone and Kenya (cf. vignettes 2 and 3), there was an explicit aim of including "all" in the country, thus enhancing greater transparency in the delivery of health services. The increasing capacity to engage with diversity enabled by the platform ecosystem was instrumental to enhancing transparency by making visible the health status of people who had not been visible before. This ability was translated to other countries when they adopted the evolving DHIS2. Other key mechanisms to link local interventions and innovations with societal consequences included the rapid growth of HISP research and education networks, the steady increase in software development capabilities, and the political legitimacy provided by the WHO and HMN partnership.

The fourth vignette is the onset of the Covid-19 pandemic. At stake, especially in the Global South, were peoples' lives and wellbeing with visible implications on the functioning of countries and economies. In such a situation, governments were made explicitly accountable through the monitoring of health status, a prerequisite for saving lives and economic security. This accountability was, by design, based on multi-sectorial data, which came from different departments and agencies and whose focus constantly changed with time. This created an unprecedented demand for a flexible, novel, and evolving recombination of multi-sectoral—including but not restricted to health—data to which DHIS2's interoperable platform ecosystem architecture was able to respond. The design and adaptation to more than 40 countries of Covid-related modules and apps were enabled by a large number of participants in the open-source DHIS2 and HISP network.

In summary, the combination of local-level digital developments and their links to societal values of equity, transparency, and accountability contributed to SDT. These links were never given, and they need to be consciously cultivated through different mechanisms, including software design and development, research and education, political advocacy, and, often, chance events (e.g., being present in a meeting when HMN was looking for a data warehouse solution). While our understanding of SDT developed through a unique history of the HISP experience, we believe that our theory lends itself to generating a set of DPs in different contexts. We discuss these DPs in the following section.

**5.2 Formulation of Design Principles (DPs)**

Abstracting from the synthesis of the learnings presented above, we articulate a set of interconnected DPs, which represent approaches to engage with SDT issues in different contexts. These DPs represent concepts that have endured the test of time (over 20 years) and space (more than 80 countries), and they arguably find relevance in a multiplicity of times and spaces. The DPs seek to address two crucial interconnected challenges: i) designing, implementing, and evolving digital solutions that are relevant for addressing local health challenges and ii) linking these local solutions to societal-level concerns to help engage with SDT.

To address the first challenge, we formulate three principles:

DP1: Start small and keep it simple: Develop locally designed generic and open solutions of high use value with minimal effort and create momentum around it to enable acceptance and scale.

DP2: Build user-led design strategies that integrate research, education, and practical system development processes.

DP3: Strengthen local ownership and capacity and build "knowledge commons" to enable the ongoing processes of scaling and sustainability.

To address the second challenge, the following three DPs are formulated:

DP4: Forge multi-faceted alliances that link successful local solutions to broader societal concerns.

DP5: Achieve full data coverage and enable the visibility of the "whole" to help intervene at the societal level.

DP6: Promote alliances through collective action around the development, interoperability, dissemination, and utilization of digital global public goods.

We elaborate on the DPs and their interconnections.

DP1 builds on key elements of the Scandinavian tradition of evolutionary participatory design (Braa & Sahay, 2012), in this context of digital transformation, together with elements from complexity theory. Although societal-level systems are chaotic and unstable (cf. Section 2), they contain one or many possible steady states around which processes converge. For example, in South Africa, the post-apartheid environment was characterized by multiple actors, old and new, the forces of history, and a desire to bring change on multiple fronts (e.g., health systems, health information systems, local autonomy, and processes of decision making) at the same time and quickly. Within this situation, the principle of essential data sets acted as a potential steady state desired by all actors, as it demonstrated the feasibility of implementing data standards thus addressing societal concern of health inequity. Given this acceptance, momentum was created, political–institutional alliances forged, and digitized essential data sets were implemented countrywide. The rapid scaling of digitized essential data sets made it possible to identify what Markus and Rowe (2021, p. 275) call "digital properties that made a difference." In our case, these are digital properties that enable SDT of the "whole" that can be processed, distributed, and analyzed (Didier, 2020) (see also DP5).

However, the design and development of digitized essential data sets were not born in a vacuum but were an outcome of ongoing processes of research, education, and practical systems development, as encapsulated in DP2. The intellectual basis for this design thinking is situated in the Scandinavian research tradition (see Section 2), which simultaneously recognizes the need for political reconfigurations to achieve multi-level and multi-stakeholder acceptance while putting user autonomy at the center of design intervention efforts. This thinking was operationalized through students' research projects hosted at the Universities of Western Cape in South Africa and Oslo in Norway, supported by ongoing incremental and agile practical systems development and implementation processes. A key to achieving success was the continuous and ongoing processes of building local ownership and capacity, as indicated in DP3, in which software development and implementation were driven by local teams comprising medical staff, software teams, activists, researchers, and ministry officials.

The bottom-up, user-led design using rapid prototyping created ownership by providing users with immediate access to their own data, which had not been possible before. This illustrates Zuboff's (1988) dilemma of technique, in which work roles and relationships change, triggering a need to leverage opportunities for learning and empowerment. A contrasting example shows that Zuboff's dilemma is real as the previously underrepresented black and colored population required a careful data representation of local-level "facts" and opened up possibilities for unintended consequences (Zuboff, 1988). In one of HISP's pilot districts outside Cape Town, the team worked on improving paper forms to capture essential data by

"ticking" a list. The old paper forms had obsolete categories from the apartheid area ("blacks," "coloreds," and "Indians"). However, health workers in dressing rooms (i.e., where you attend to wounds, plastering, and post-surgery) used these parts of the form to report other activities they deemed important. Removing the categories used to report the "true" representation of their work was seen as making their work invisible, and the health workers argued that it would put their jobs at risk. Thus, the new essential data form was not implemented.

Although connected with each other, DP1–DP3 need to explicitly support the DPs aimed at societal-level transformations (DP4–DP6). The mobilization around the digitized essential data set in South Africa was elsewhere mimicked in different forms through different technical artifacts, such as data warehouse and platform, at different temporal and spatial points in history. Nevertheless, they all played a similar role: enabling political–institutional alliances in which momentum for societal change could be mobilized and strengthened (DP4). The fully open-source DHIS2 data warehouse capabilities served as the key vehicle for gaining political credibility at the national (e.g., Kenya) and global (e.g., WHO and HMN) levels. This not only served as important entry points but also helped enroll multiple other actors (e.g., WAHO, National Ministries of Health, and other universities) that helped set in motion various other processes of national and societal significance. Although specific issues of societal significance will vary in different contexts, an enduring concern is transparency, which reflects how well the state is able to serve the health needs of the entire population (i.e., equity). Bringing "visibility of the whole" to the fore becomes an important principle (DP5) of societal consequences across countries, particularly in the South. The online cloud-based data warehouse approach enabled the capability of digitized data of the whole to be processed, analyzed, and disseminated. However, with online access and increased digital maturity, users and countries wanted to further explore the "whole." As a director in GAVI once said, "I am not interested in the children in the immunization register, as they have their vaccines. I am interested in the children who are not there." In Laos, a design effort to represent all children for immunization, whether represented in the register or not, went through two steps (Byrne et al., 2022). First, a case-based immunization register in which each child was linked to the home village was developed and implemented on DHIS2. Second, to identify the children not represented in the register, other data sources (e.g., population census, health data on reported births) were applied to estimate the number of infants due for immunization in each village. By comparing the number of represented children in these villages, the number of not-immunized children in each village could be estimated or indirectly represented. This digital representation of the non-represented—including the excluded (Castells, 2011)—was achieved indirectly through the combination of automating and informating (Zuboff 1988), providing new combinations of information and new potentials for use (Didier, 2020). Therefore, the visibility of the whole needs to be supported by appropriate technical solutions (DP1), interoperability should enable the sharing of data across data sources (DP6), effective networks and alliances must be in place (DP4), and there must be ownership and capacity in digital systems to continuously ensure scale and sustainability in countries.

An overarching principle to ensure DP1–5 can effectively function concerns the principle of digital public goods (cf. Section 2), which normatively seeks to ensure the non-rivalrous and non-exclusive delivery of goods (in our case, digital innovations) to whole populations (DP6). In the case of DHIS2, the open source capabilities of the digital public goods concept have been a key digital property of making a difference by enabling network-based design and "partitioned" innovations in open source networks (Tiwana, 2013) and by the creation of real ownership of the software platform in countries, as they "own" the source code and are designing their own systems. For instance, during the Covid-19 period, Sri Lanka, Rwanda and Mozambique established platform ecosystems in their digital responses to the pandemic.

This capacity of interoperable platform ecosystems illustrates a "digital property that made a difference" (cf. Markus & Rowe, 2021).

DP6 captures how the DHIS2 case enable networking dynamics between and within demand-side and supply-side users in line with Gawer's (2014) platform ecosystem. Sahay et al. (2013) pointed out the different "distortions" in countries that constrain the achievement of the ideal of public goods, such as the political environment, capacity, and infrastructure, which need to be consciously engaged with. Castells' (2011) ideas of counter-network emphasize that joining the network society is not trivial but requires long-term, intensive, and innovative efforts for enabling the marginalized to obtain the benefits of DTs.

In summary, we propose that DP1–6 serve as the seed for a design theory on SDT. For analytical purposes, we divided them into two categories related to local-level design solutions and designed mechanisms to link them to matters of societal consequences. We argue that these principles are inextricably interconnected and that the failure of one will significantly constrain the achievement of SDT. In the following section, we provide an overview of our theoretical contributions.

**5.3 Theoretical Implications of our SDT Theory**

Our DPs are formulated based on a longitudinal analysis of flows and processes that have unfolded over an extended period of time and in multiple countries. As we have noted (cf. Section 2.2), the open source-based platform around which HISP is organized exhibit (digital global) public good characteristics. However, with the multiplicity of flows, including software development, research and education, political advocacy, and networking, this amounts to governing a "knowledge commons" that vary significantly over space and time (Frischmann et al., 2014). DPs, by their very nature, can appear to be reductionist, as the complexity involved cannot be reduced to simple maxims. While acknowledging this risk, we discuss three implications related to salient characteristics of our theory.

First, there is a general paucity of IS research on longitudinal or historically informed studies that also apply to DT. There are precious few studies of DT (with or without this explicit label) on change processes that span a number of years but not decades (exceptions include Ribes and Finholt (2009)). Clearly, from our analysis, a longitudinal perspective is the only manner to which the constantly changing technological artifact responds and feeds into an equally evolving organizational, institutional, and political context. An interesting and relevant theoretical formulation is Williams and Pollock's (2012) notion of a "biography" of artifact. Consistent with our analysis of theorizing, a biographical lens underscores the "becoming" of technology. As Baygi et al. (2021, p. 444) noted, "[e]nclosed in every 'actor' or 'entity' there is a story of how it became what we now take it to be, and where it is heading."

Second, with an emphasis on processes unfolding over time, we are partial to the perspective that privileges flows over actors/structures, as suggested by Baygi et al. (2021, p. 424), "to foreground how contingent, unpredictable, and sometimes seemingly insignificant confluences of heterogenous flows of action explain the conditionalities and directionalities of the course of (trans)formation." Grappling with irreducible uncertainty, we advocate systematically postponing design decisions when possible by investing in options (Tiwana, 2012), an approach consistent with a knowledge commons perspective (Frischmann, 2012; Frischmann et al., 2014). Our approach highlights long-term and yet multiple timeframes, which help discern the shifts in the ebbs and flows of changing political sentiments, policies, value propositions, and ideologies. This entails viewing these as processual rather than imposed, as "the idea of following a flow of events raises questions about how to conceptualize policy and its movement" (Shore & Wright, 2011, p. 12).

SDT is by definition characterized by flows at multiple levels, including global, societal, health system (in our case), health facility, and community, and a focus only on one level encompasses the complexity and interconnected nature of SDT (Folke et al., 2010). Using Baygi et al.'s (2021) metaphor of a rope, different flows of the digital, the health system, and the wider context are joined together in varying forms of correspondences, enabling transformational processes at different scales, both in the whole and in the parts. SDT comprises a multiplicity of flows, each with different temporal trajectories or rhythms, including hard-to-change competing and dominant flows.

Therefore, we extend Baygi et al.'s (2021) conceptualization of correspondence primarily at one level by emphasizing a multi-level perspective that creates multiple possibilities for action at different levels. An example of this is DP5 concerning the "data of the whole," which consists of representations of health (and other domains) conditions ranging from the individual to the national level.

Correspondences of flows enable varying action possibilities at different intersecting levels. Design approaches formulated in DP1 create data representing different flows of activities, and health services, when aggregated, help represent the "whole." Only a multi-level and interconnected perspective can help address different concerns of equity across geography, racial groups, and communities (Didier, 2020). While top-level administrators can gather evidence on building interventions to address equity, low-level health workers can better represent their work, and system developers can evolve more effective data collection tools and artifacts to better represent issues of societal concern. Correspondences are also developed laterally. For example, a district officer's ability to compare local performance to that of others may enable empowering influences (Braa & Hedberg, 2002) and highlight Zuboff's (1988) dilemma of technique, representing new skills required to do such analysis.

Third, our theorizing of SDT problematizes the notion of "value," which is significantly different from the market- or consumer-oriented ways of understanding value commonly emphasized in IS in the DT literature (Barthel, 2021, Reis et al., 2018, Vial ,2019, Wessel et al., 2020). By contrast, SDT spans multiple levels and engages with multiple value systems. Values stem from different sources, including the underlying forms of representation and how they are used (Burton-Jones, 2014, Leonardi, 2012). Informating processes taking place at the individual and group levels can have varied and often unintended effects. For dressing room health workers in South Africa, the "textualization" or "datafication" (Lycett, 2013) of their work (filling of essential and not all data representing their work) led them to believe that important aspects of their everyday job were made invisible. Similarly, for the information officers who could now present facility-wise comparisons of reports to the district management, informating processes were considered to add value. Therefore, value is a multi-faceted concept (Boltanski & Thévenot, 2006) tied to different dilemmas of knowledge (Zuboff 1988), creating uneven and unintended effects at different levels and for varying groups in the health system. Such differential effects are not visible with a unitary and snapshot perspective, which our SDT perspective seeks to highlight.

**5.4 Practical Implications**

We discuss practical implications of the SDT for three target audiences illustrated by two examples The audiences are practitioners in specialized domains across sectors, policy makers and systems designers/ developers. The implications for the latter amounts to how DPs are appropriated to respond to demands from the former two audiences.

Designing systems for specialized domains, a first step is to operationalize the WHO digital health package approach. This starts with identifying the list of essential indicators with guidelines for how they are to be configured, analysed and presented in dashboards with charts, maps and tables (Poppe et al., 2021). This is a practical way to develop digitally enabled knowledge commons and community-driven development across sectors (DP1, DP3). To illustrate this approach, a program under the Ministry of Infrastructure in Mozambique for monitoring water, sanitation and hygiene in all villages in the country, focused on data analysis and dashboards available at their public web site[23]. Water sources and latrines are categorized and monitored according to safety and hygiene standards and villages are certified according to a score (i.e., an indicator). When a water pump is not working, villagers update the system by sending a SMS text message. The scaling and flexibility such digital domain packages are illustrated by a current effort to design a similar system in Angola, now under the Ministry of Education, to monitor water and sanitation in schools (DP3, DP4, DP6). However, further development of user-led design strategies (DP2) is required to develop, share and maintain digital domain packages.

Societal level transformations necessarily need to address the policy makers, who have the resources and authority to address changes at scale. The vignettes have all described design strategies addressing country policy levels, from the designing for equity in South Africa and accountability in Kenya by aiming for coverage and the "whole" (DP5), to optimizing in-country capacity (DP2, DP3) and open source software (DP6) to enable national control during the pandemic. Accountability illustrates a policy area where our SDT may remain useful and where systems design increasingly depends on appropriate approaches to standardization (minimum data sets) and interoperability. We exemplify. Building on in-country capacity and experience with DHIS2 (DP3, DP6), Rwanda is designing a new system monitoring their Key Performance Indicators (KPI) from all 16 government sectors (including energy, agriculture, and health) and all districts, i.e., literary the "whole" (DP5). Discussing design strategies with the project team, a lack of standards across sectors making interoperability difficult was identified as a main challenge. Particularly important design principles were identified as: Start small and keep it simple in terms of scope and data sets and to focus on developing early useful solutions to convince stakeholders to follow suit (DP1).

---

[23] https://sinas.dhis2.org/portal/

## 6 Conclusion

Our design theory for SDT is based on a truly unique experience that is difficult, not to say impossible, to replicate, given its institutional, political, and technological embedded circumstances. This triggers the reasonable question of what, if any, part of our theory/ model has relevance and significance in other circumstances. Given the uniqueness of the experience, are the principles that make up our model generalizable or transferrable to other contexts? We argue that the boundary conditions for our theory are significantly wider than the idiosyncratic nature of the above narrative may suggest.

First, we described HISP as operating in the Global South as if such a thing exists. However, the label is inaccurate because it could be seen to perpetuate a view of the Global South as homogenous or "flat" (Ritzer, 2013), which, of course, is a pure myth or fantasy. Space limitations prohibit us from detailing the plasticity of our principles when applied as appropriate to the dramatic differences on so many levels and in so many ways across the more than 80 countries of the HISP. As an indication, DP2, which involves the strategy for implementation, covers radically different approaches: (i) negotiating access to a selected few pilot sites (as illustrated in South Africa), (ii) securing top-level clearance in centralized institutional settings (as illustrated in Rwanda), and (iii) a regionalized approach applicable in a federated institutional setting (as illustrated in WAHO in West Africa).

Second, our analysis covers HISP's engagement in one domain: healthcare. This hides the fact that DHIS2 has traveled to a number of other domains, thus giving credibility to the transferability of our design theory across a variety of domains. A significant ongoing initiative is to adopt our DPs in the educational sector. Well underway in Gambia and Uganda, our model—tweaked and appropriated—acts as a forceful guiding principle. Similarly, DHIS2 has been appropriated to logistics/ supply-networks, vaccine cold chain monitoring, and human relations. Through HISP India, the United Nation's Food and Agriculture Organization has designed and developed a Food Security and Nutrition Information System for the Ministry of Food in Bangladesh. While writing these lines, one of the authors is part of an African Union effort to address occupational health in the mines in Africa using the DHIS2 to track safety-related incidents. It involves the ministries of mines, labor and health, thus illustrating the transfer to another domain by HISP and DHIS2.

Third, the products and processes embedded in our design theory have traveled to the North. Norway, one of the world's most affluent societies, struggled with the rampant Covid-19 pandemic. Slow, labor-demanding, and often paper-based, Norway's monitoring system had a hard time keeping up with the pace and mutations during the pandemic. The Norwegian Institute of Public Health, which corresponds to the CDC in the United States, collaborated with HISP to repurpose DHIS2 modules to increase the efficiency and speed of disease surveillance in a number of Norwegian municipalities.

In sum, the outlined experiences of the transferability of our design theory across significant variations in terms of institutional fabric and domains and across the North/ South dichotomy speak loudly of the considerable scope and relevance of our design theory. As Nunes (2021, p.12) argued, when discussing societal problems "of that magnitude and complexity, the most plausible alternative seems to be some kind of distributed action combining organization at different levels and scales," we expect our design theory to be especially relevant to social movements, political campaigns and/or ideologically motivated networks. We welcome and encourage research program to explore more comprehensively and systematically the limits and possibilities of its applicability.